\documentclass[mathleft
]{an}
\usepackage{graphicx}
\usepackage{times}
\usepackage{natbib}
\bibpunct{(}{)}{;}{a}{}{,}
\begin{document}

\Pagespan{789}{}
\Yearpublication{2016}%
\Yearsubmission{2015}%
\Month{11}%
\Volume{999}%
\Issue{88}%

\title{The Future of X-ray Reverberation from AGN}

\author{A.C. Fabian\inst{1}\fnmsep\thanks{Corresponding author:
  \email{acf@ast.cam.aac.uk}\newline} W.N. Alston\inst{1},
  E.M. Cackett\inst{2}, E. Kara\inst{3}, P. Uttley\inst{4} \and 
D.R.Wilkins\inst{5}}
\titlerunning{Future of X-ray Reverberation}
\authorrunning{A.C. Fabian et al.     }
\institute{1. Institute of Astronomy, University of Cambridge, Madingley Road,
Cambridge CB3 0HA, United Kingdom\\
2. Department of Physics\& Astronomy. Wayne State University, 666
W. Hancock St., Detroit, Mi 48201, USA\\
3. Department of Astronomy, University of Maryland, College Park, MD
20742-2421, USA\\
4. Anton Pannekoek Institute for Astronomy, University of Amsterdam,
Science Park 904, 1098 XH Amsterdam, The Netherlands\\
5. Physics Department, Stanford University, 382 Via Pueblo Mall,
Stanford, CA 94305-4060\\
}

\received{Oct 2016}
\accepted{Oct 2016}
\publonline{later}


\abstract{XMM-Newton is capable of making a transformational advance
  in our understanding of how luminous accreting black holes work, by
  dedicating about 10\% of future observing time to long observations,
of order Megaseconds, to X-ray variable Active Galactic Nuclei (AGN) research. 
This would
enable reverberation studies, already a commonplace feature of AGN, 
to proceed to the
next level and follow the behaviour of the powerful dynamic
corona. Such a dedicated legacy programme can only be carried out with XMM-Newton.}

\maketitle

\section{Introduction} The current picture of the inner regions of a
luminous Active Galactic Nucleus (AGN) shows an accretion disc
extending down to the Innermost Stable Circular Orbit (ISCO) around a
supermassive black hole. Magnetic fields extending upward from the
disc power a compact corona \citep{Galeev79,Haardt93}.  
Energetic electrons in the corona
Compton scatter soft thermal photons from the accretion disc to
produce a power-law X-ray continuum. This emission in turn irradiates
the disc to produce fluorescent and other lines with backscattered
continuum X-rays, together known as the reflection spectrum
\citep{george91,fabian10}.  If the
coronal power varies, we see changes in both the power-law continuum
and in the reflection spectrum, but with a delay between them due to
the longer light path taken by reflection: this is the process of
reverberation \citep{Uttley14}.

\begin{figure*}
\includegraphics[width=0.98\textwidth]{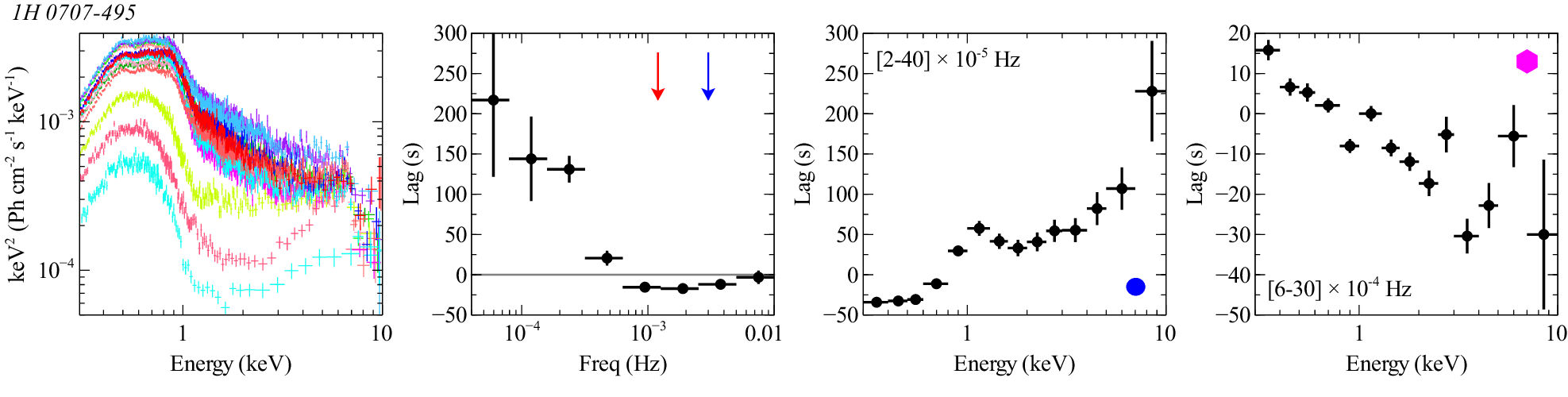}
\includegraphics[width=0.98\textwidth]{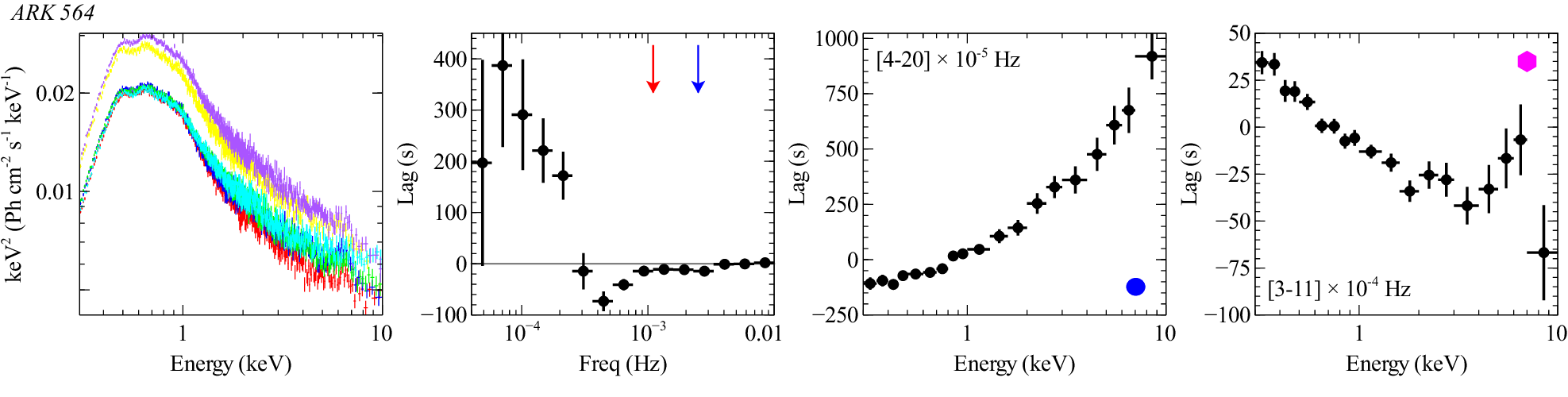}
\includegraphics[width=0.98\textwidth]{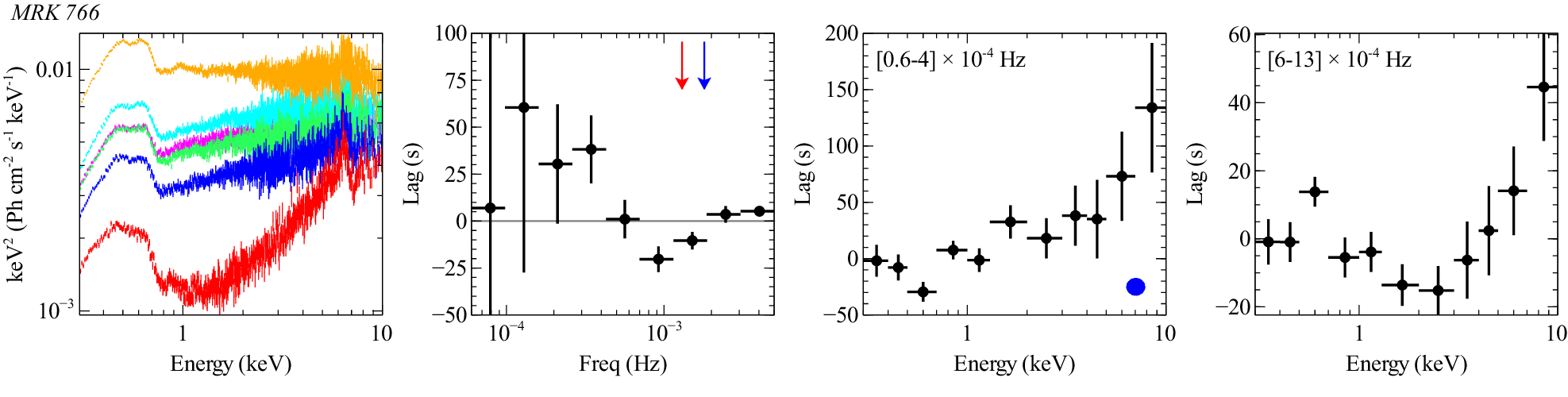}
\caption{From left to right: Energy spectrum, lag-frequency spectrum, low
  frequency lag-energy spectrum, high frequency lag-energy spectrum for (top to
  bottom) 1H0707-495, Ark 564, MRK766; \cite{Kara16}. Note in Ark564
  that the low frequency lag energy spectrum is featureless, whereas
  the high frequency one reveals an iron line, showing that it is due
  to reflection. Mrk766 has no obvious iron line in its high frequency lag
  energy spectrum. Its behaviour resembles that of MCG--6-60-15.   
}
\label{label1}
\end{figure*}

The observer sees the sum of the coronal power-law and the delayed
reflection, which also has a continuum component. They can be
separated since the reflection spectrum differs in shape from the
coronal power-law, due to atomic emission and absorption processes as
well as Compton recoil (for analysis techniques see
\cite{Uttley14}). It generally appears as a soft excess at low
energies below 1 keV, a broad iron line from about 4--7 keV and a
Compton hump above 10 keV. This means that lightcurves made in soft
X-rays (0.3--1 keV)and the iron-K band lag behind those made in bands
dominated by the coronal power-law emission. Using many different bands, a full
reverberation energy spectrum has been made in many of the brightest X-ray
variable AGN (Fig.~1), revealing the broad iron line characteristic of inner
disc reflection which confirms the above picture. The lag timescales
indicate that the separation of the corona from the disc is often
small, about 10 gravitational radii ($10r_{\rm g}=10 GM/c^2$) or less
\citep{demarco13,cackett14,Emmanoulopoulos14,Chainakun15}.

The combination of reflection and reverberation means that we have a
powerful tool with which to explore luminous accretion onto black
holes.  They can help to solve outstanding questions about the origin,
shape and activity of the corona. How is the corona powered, how does
it change in size or location as the power increases or decreases? Is
the corona related to  jets and/or winds from AGN? How well can we 
determine the radius of the ISCO and thus the
spin of the black hole? Does the disc truncate as its power or the
mass accretion rate drops? Reverberation/reflection studies have the
potential to transform our understanding of the inner regions of
accreting black holes.

Rapid X-ray reverberation in AGN was discovered with XMM-Newton \citep{Fabian09}
and most results have since been obtained with XMM-Newton, together with a
handful of NUSTAR contributions which have revealed the reverberation
of the Compton hump \citep{Zoghbi14}.  About 50 per cent of a sample of 43
variable AGN observed with XMM-Newton show high frequency, iron K
reverberation which indicates that reflection is involved
\cite[Fig.~1]{Kara16}. The key requirements for such studies are lots
of photons (defining ``variable counts'' to be total counts times 
fractional variability, $>7,000$ are needed, Fig.~2) and
long observing times ($>40$~ks \cite{Kara16}).

\begin{figure}
\includegraphics[width=0.95\columnwidth]{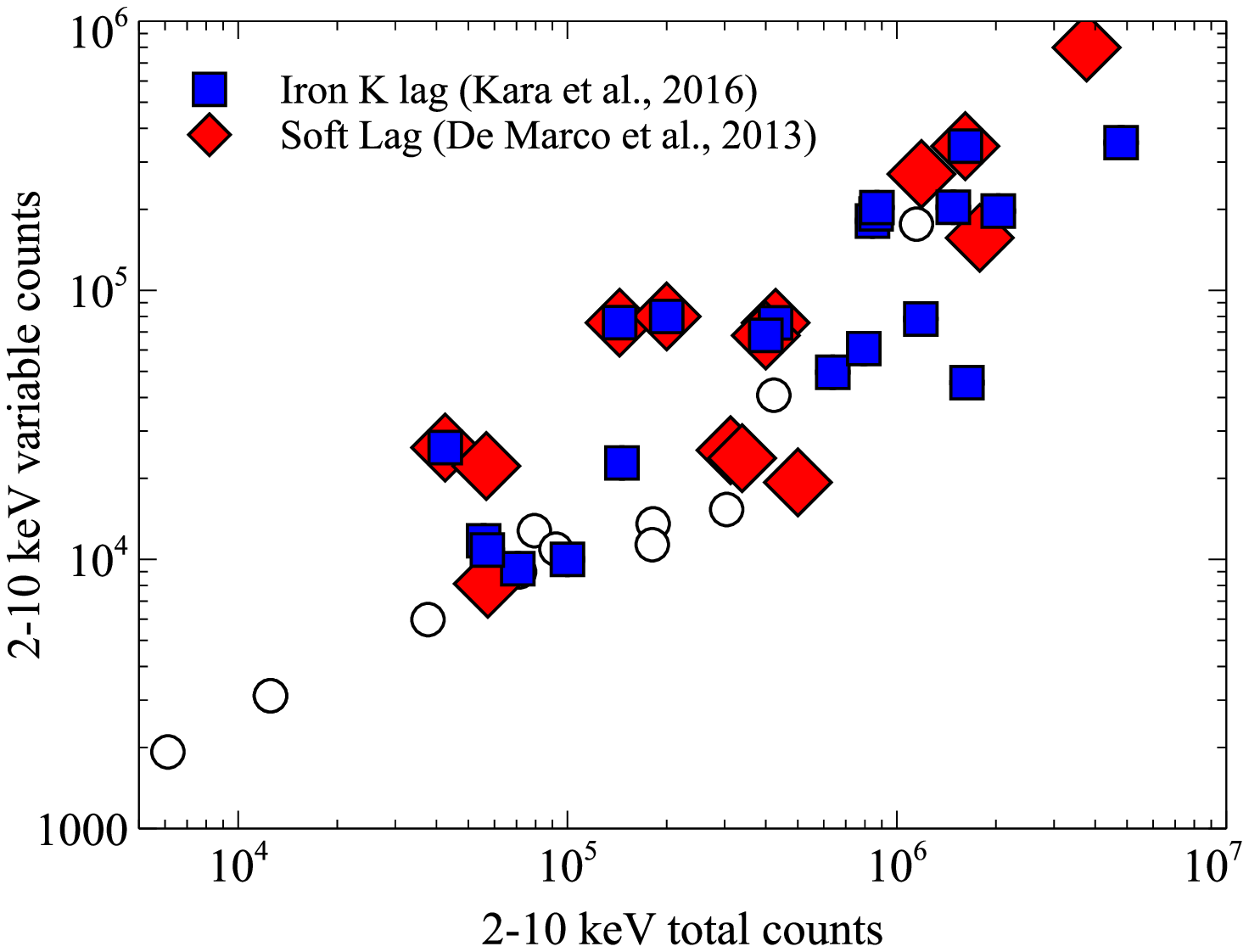}
\caption{Variable counts in the 2-10 keV band plotted against total
  counts for 43 variable AGN (adapted from \cite{Kara16}). Open circles
  represent non-detections. 80\% of AGN with more than 7,000 variable
  counts (total counts times fractional variability) show
  reverberation lags. 
Sources with soft lags \citep{demarco13} are marked by a red diamond.
}
\label{label1}
\end{figure}

\begin{figure}
\includegraphics[width=0.95\columnwidth]{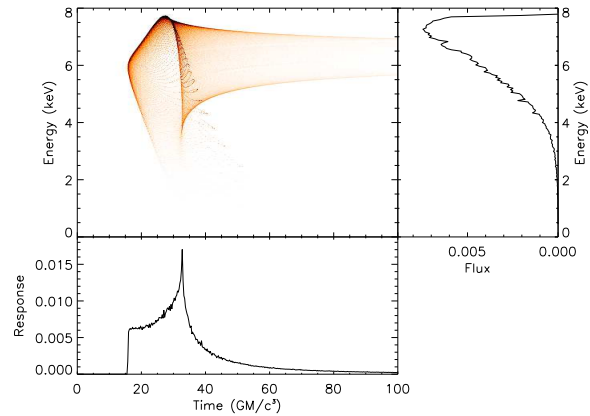}
\caption{ Theoretical energy -- time-lag diagram from single
  irradiation flash from a point-like corona at a height $10$~r$_{\rm
    g}$ above the centre of an accretion disc above a rapidly spinning
  black hole. The disc is inclined at 60 degrees. What is seen is the
  development of the reverberation of a 6.4 keV iron fluorescent
  line. The largest energy spread comes from the smallest radii. The
  projection onto the energy axis shows the broad iron line (to the
  right) and onto the time axis the impulse response function (for all
  energies). See \citep{cackett14,campana95,reynolds99}.
}
\label{label1}
\end{figure}

\begin{figure}
\includegraphics[width=0.7\columnwidth,angle=-90]{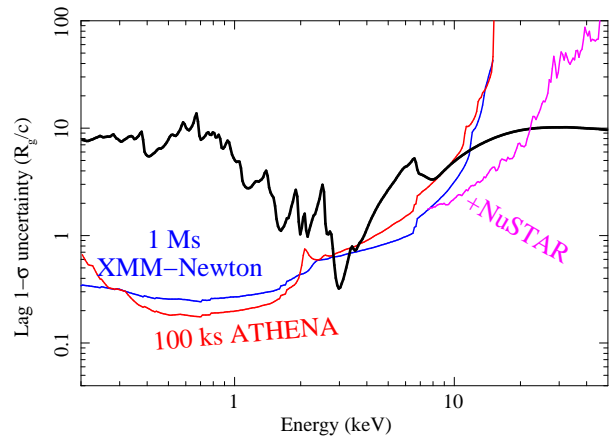}
\caption{Simulation showing achievable sensitivity in units of
  $r_{\rm g}/c$, adapted from \citep{Uttley14}. The black line is a
  'sample' 
lag-energy spectrum (for a $10^6$ M$_{\rm \odot}$  AGN).  The assumed 
2--10 keV flux is $4\times 10^{-11}$  cgs (for AGN
  everything scales with $\sqrt{\rm flux}$ so it is easy to see what is
  needed for other flux levels).  The lag errors just scale with
  $1/\sqrt{\rm exposure\,time}$. Note that whilst a 100 ks ATHENA
  exposure can statistically match a 1 Ms XMM-Newton exposure on high
  frequency lags (modelled here) it cannot reach the lowest frequency 
propagation  lags ($\ll 10^{-4}$). Similarly, the long XMM-Newton
exposures yield an average result 
  whereas a shorter ATHENA exposure can follow the changes in a 
  corona on 100 ks timescales.     }
\label{label1}
\end{figure}

Even with a few 100ks of exposure (ie 2--3 XMM-Newton orbits), the typical
current reverberation data quality allows identification of only a
characteristic reverberation time which translates roughly to the
height of the corona from the disc. This typically is about
$3-10 r_{\rm g}$. The archival data have now been thoroughly explored
\citep{demarco13, Kara16}. As shown in Fig.~3, the
reverberation signal is richer than this and it should be noted that
the corona is also likely to be a dynamic not static
phenomenon. Pushing the subject forward so that we can explore and map
the inner regions of the X-ray emission region around luminous
variable accreting black holes requires a leap of 3 to 10 in observing
time for a significant sample of AGN (Fig.~4). This means exposures of 300 ks
to 1Ms.

\begin{figure}
\includegraphics[width=0.95\columnwidth]{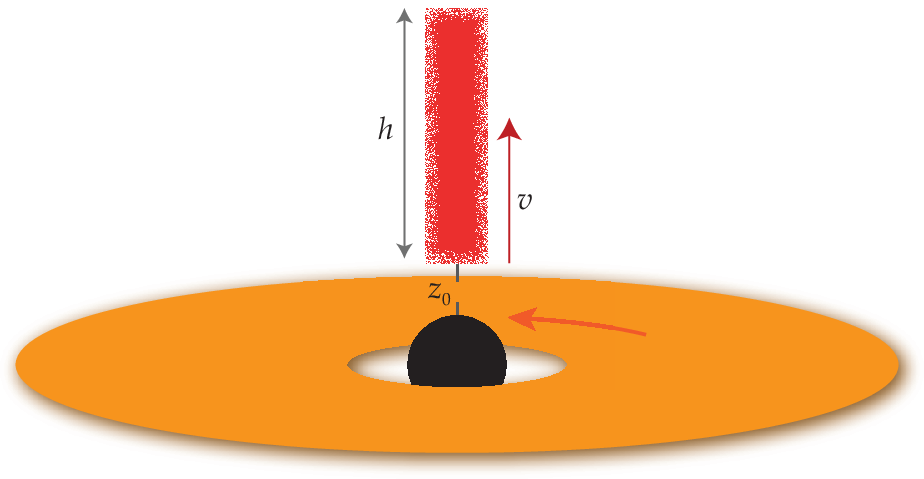}
\includegraphics[width=0.95\columnwidth]{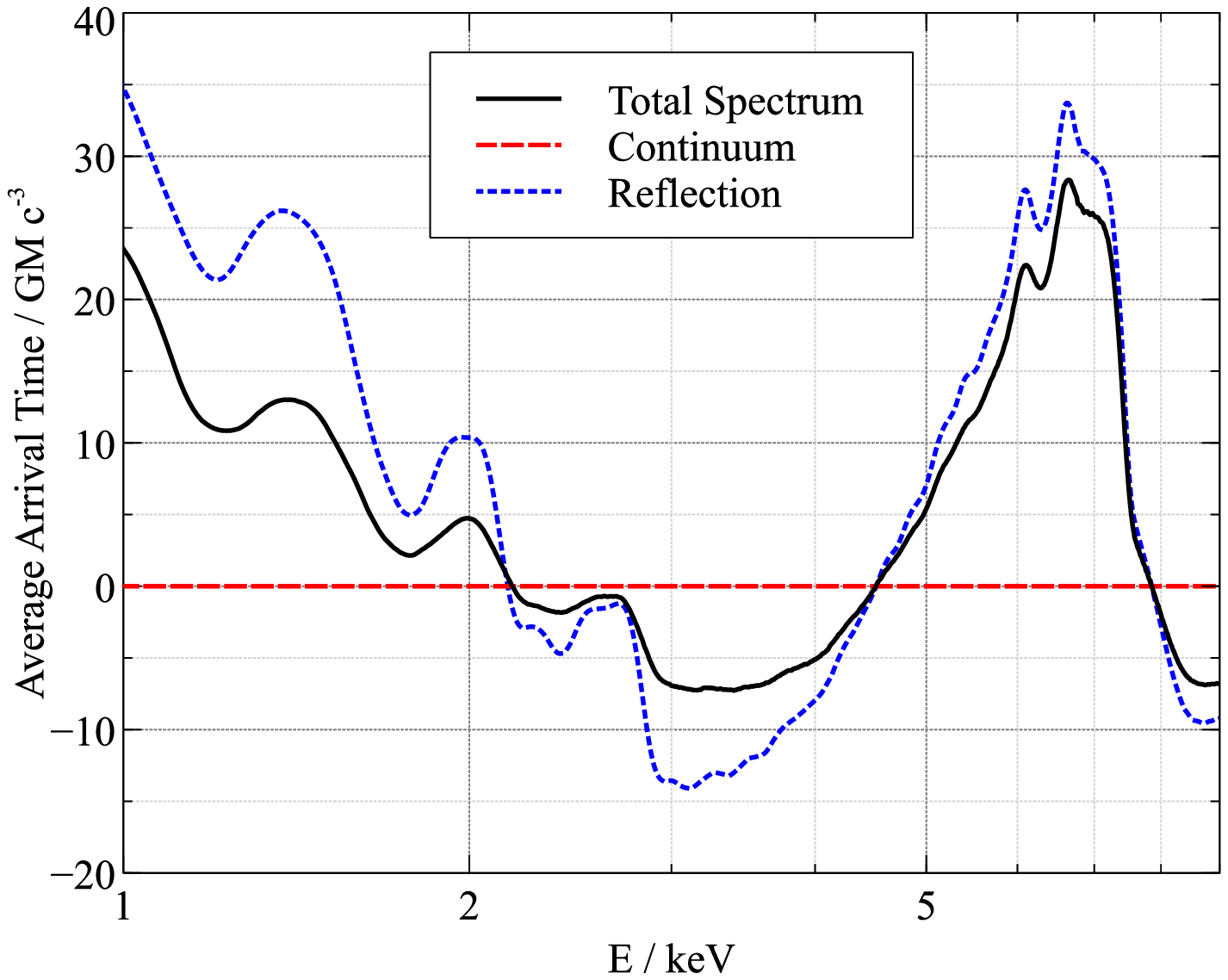}
\includegraphics[width=0.95\columnwidth]{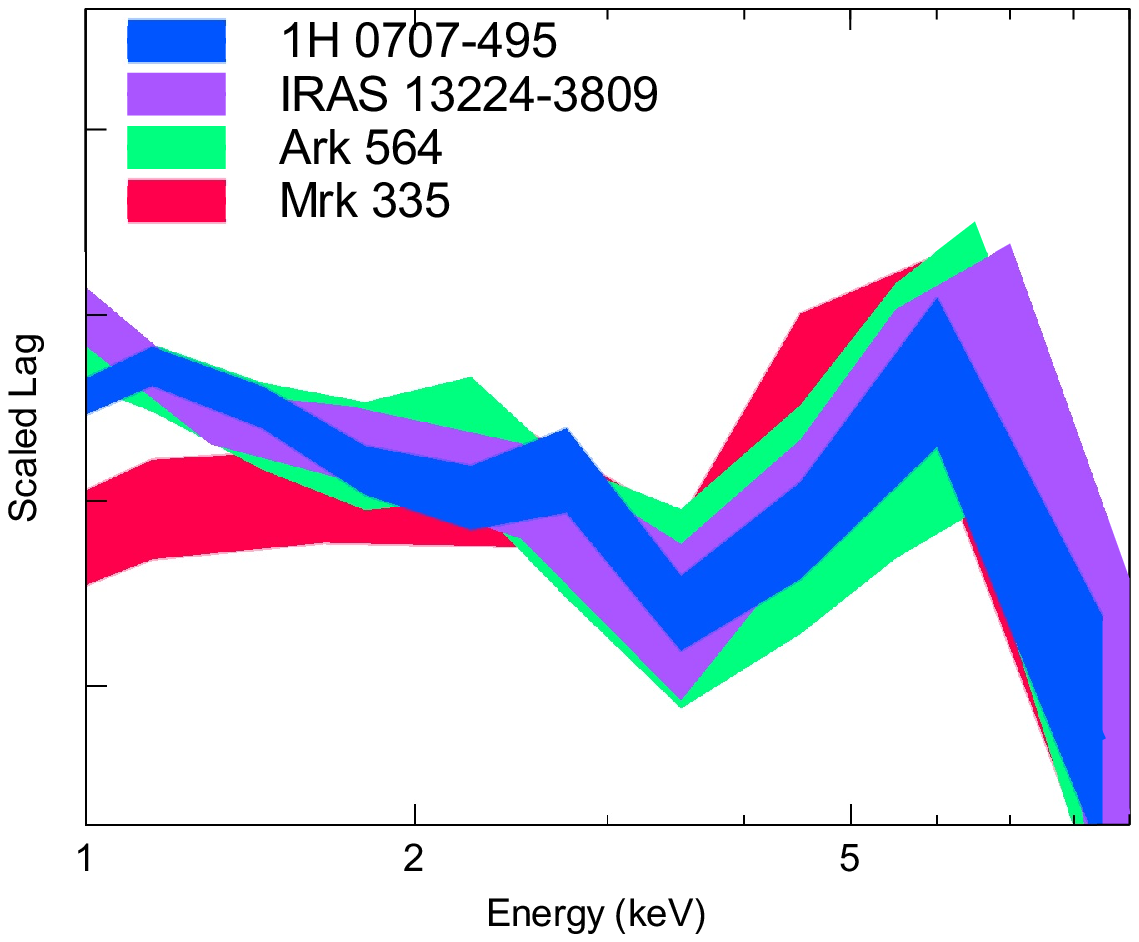}
\caption{Top: Schematic model of vertically extended corona in which
  the matter flows upward at velocity $v$ \citep{Wilkins16}. Centre: 
Lag energy spectrum resulting from this model
  when $v$ increases upward with height $h$. Lower: Lag energy
  spectrum for 4 objects in which the dip at 3--4 keV matches that
  seen in the centre panel.   
}
\label{label1}
\end{figure}

Studying a sample drawn from the  33 objects with more than 5,000 variable 
counts in Fig.~2
would be an excellent start. As well as examining in more detail the
known lag sources we need to explore why some sources do not yet show
lags despite reasonable exposures.

At its current observing efficiency, XMM-Newton obtains data for about 18 Ms
per year, so over the next 5 years, the total integrated exposure
should be 90 Ms. If 10 per cent of this observing time is dedicated to
long (Ms) AGN exposures then a sample of about 10--20 objects can be
studied in depth.  What this represents is a committed
reverberation/reflection study to understand how the central engine in
the most luminous persistent objects in the Universe operates, namely
the accreting supermassive black holes.  It is a legacy study that
cannot be carried out at any other wavelength.
 
\section{Some Details}

A pulse of emission from a point-like corona reverberates off the disk
in the manner shown in Fig.~3. The result changes with inclination. 
The reverberation signal in both temporal frequency and energy
contains valuable information on the size and geometry of the inner
accretion flow and the corona. Most current work has only captured the
gross aspects of the the reverberation process, yielding a
characteristic time associated with the bulk of the return signal.

This constitutes the simple lamp-post model in which the corona is a
point source on the rotation axis of the accretion disc. It makes the
situation computable and minimises the number of free parameters. Note
that the computations need to be carried out in the Kerr metric, with
strong light bending and energy shifts. In practice of course the
corona is extended, either radially and/or vertically as discussed by
\cite{Wilkins16}.  It probably changes size.  If for example it is
extended vertically (Fig.~5) and its length $h$ varies, then strong 
light bending
means that the upper end may dominate the observed continuum while
the lower end dominates the reflection. This could be the situation in
MCG--6-30-15 and Mrk766 which both show soft lags (0.3--1 vs 1--4
keV), but the continuum
and iron line are not well correlated and no Fe-K reverberation is
seen as yet.

The coronal matter may be outflowing through the corona leading to the
continuum being beamed away from the disc at mildly relativistic
velocity.  This situation has been modelled by \citep{Beloborodov99} and
can plausibly explain the low reflection fraction seen in some
objects. The corona may evolve from being static to outflowing in an
object, leading to an apparent continuum outburst as seen in Mkn 335
\citep[Fig. 4]{Wilkins15, Wilkins16}. It is also plausible that the corona is 
corotating with the disc beneath it and beaming coronal radiation along
the disc. In other words, the corona is likely to be dynamic, rather
than a fixed object. To understand it we must be prepared
to follow its changing nature.

All of the effects discussed above occur at high temporal frequency so
we are dealing with variations on the light-crossing time of the inner
region (frequency $f<c/50r_{\rm g}$ which is typically above $2\times
10^{-4}{\rm Hz}$ for black holes of mass $10^6 -10^7\,{\rm
  M_{\odot}}$). Below that
frequency ($f<2\times 10^{-4}{\rm Hz}$) it is common to see {\it hard} lags
in which the 1--4 keV band lags the soft X-ray band of 0.3--1
keV. These are not due to reverberation but appear to be intrinsic to
the process powering the corona. Such low frequency hard lags are
common in accreting black hole systems and have been seen for a long
time, being first seen in the Black Hole Binary Cyg X-1
\citep{Miyamoto89}. Perhaps the best description of them is as
fluctuations created and propagating through the accretion flow
\citep{Kotov01}.

Low-frequency, hard lags typically have a log-linear energy spectrum
which does not show reflection features, unlike the high-frequency
soft lags.  They may however have a high frequency contribution which
interferes with the detection of the true reverberation signal in some
objects. They may be modelled and corrected for if we understood their
origin better. 

Finally, there can be variable absorption along the line of
sight. This can be modelled and corrected as in NGC1365 \citep{walton14},
revealing the reflection spectrum underneath.  It represents a source
of additive noise to the variability process. Low frequency time lags
can be affected by distant absorption (e.g. NGC1365; \citep{Kara15}),
therefore time lag studies can place independent constraints on
complex absorption \citep{Silva16}. 

\section{Summary}
A dedicated 5 year, 9\,Ms, XMM-Newton legacy AGN reverberation programme 
is expected to 
transform our understanding of the innermost energy-generation region
(the central engine) of luminous accretion onto black holes. It will
enable us to go beyond the simple detection of a characteristic
timescale and reveal the corona in dynamic detail, its interaction
with the accretion flow and the inner accretion disc
immediately around the black hole.

\section{Acknowledgments}
Thanks to the members of the reverberation study group of the X-ray
Spectral Timing Revolution Workshop at the Lorentz Center in Leiden
this February for stimulating discussions.  ACF thanks the European
Research Commission for Advanced Grant FEEDBACK 340442. WNA and ACF
acknowledge grant StrongGravity1351222 from the EU 7th Framework
Programme.

\bibliographystyle{an}
\bibliography{fabiana_lib}

\end{document}